\catcode`\@=11					



\font\fiverm=cmr5				
\font\fivemi=cmmi5				
\font\fivesy=cmsy5				
\font\fivebf=cmbx5				

\skewchar\fivemi='177
\skewchar\fivesy='60


\font\sixrm=cmr6				
\font\sixi=cmmi6				
\font\sixsy=cmsy6				
\font\sixbf=cmbx6				

\skewchar\sixi='177
\skewchar\sixsy='60


\font\sevenrm=cmr7				
\font\seveni=cmmi7				
\font\sevensy=cmsy7				
\font\sevenit=cmti7				
\font\sevenbf=cmbx7				

\skewchar\seveni='177
\skewchar\sevensy='60


\font\eightrm=cmr8				
\font\eighti=cmmi8				
\font\eightsy=cmsy8				
\font\eightit=cmti8				
\font\eightbf=cmbx8				

\skewchar\eighti='177
\skewchar\eightsy='60


\font\ninei=cmmi9
\font\ninesy=cmsy9

\skewchar\ninei='177
\skewchar\ninesy='60


\font\tenrm=cmr10				
\font\teni=cmmi10				
\font\tensy=cmsy10				
\font\tenex=cmex10				
\font\tenit=cmti10				
\font\tensl=cmsl10				
\font\tenbf=cmbx10				
\font\tentt=cmtt10				
\font\tenss=cmss10				
\font\tensc=cmcsc10				
\font\tenbi=cmmib10				

\skewchar\teni='177
\skewchar\tenbi='177
\skewchar\tensy='60

\def\tenpoint{\ifmmode\err@badsizechange\else
	\textfont0=\tenrm \scriptfont0=\sevenrm \scriptscriptfont0=\fiverm
	\textfont1=\teni  \scriptfont1=\seveni  \scriptscriptfont1=\fivemi
	\textfont2=\tensy \scriptfont2=\sevensy \scriptscriptfont2=\fivesy
	\textfont3=\tenex \scriptfont3=\tenex   \scriptscriptfont3=\tenex
	\textfont4=\tenit \scriptfont4=\sevenit \scriptscriptfont4=\sevenit
	\textfont5=\tensl
	\textfont6=\tenbf \scriptfont6=\sevenbf \scriptscriptfont6=\fivebf
	\textfont7=\tentt
	\textfont8=\tenbi \scriptfont8=\seveni  \scriptscriptfont8=\fivemi
	\def\rm{\tenrm\fam=0 }%
	\def\it{\tenit\fam=4 }%
	\def\sl{\tensl\fam=5 }%
	\def\bf{\tenbf\fam=6 }%
	\def\tt{\tentt\fam=7 }%
	\def\ss{\tenss}%
	\def\sc{\tensc}%
	\def\bmit{\fam=8 }%
	\rm\setparameters\setbaselines\fi}


\font\twelverm=cmr12				
\font\twelvei=cmmi12				
\font\twelvesy=cmsy10	scaled\magstep1		
\font\twelveex=cmex10	scaled\magstep1		
\font\twelveit=cmti12				
\font\twelvesl=cmsl12				
\font\twelvebf=cmbx12				
\font\twelvett=cmtt12				
\font\twelvess=cmss12				
\font\twelvesc=cmcsc10	scaled\magstep1		
\font\twelvebi=cmmib10	scaled\magstep1		

\skewchar\twelvei='177
\skewchar\twelvebi='177
\skewchar\twelvesy='60

\def\twelvepoint{\ifmmode\err@badsizechange\else
	\textfont0=\twelverm \scriptfont0=\eightrm \scriptscriptfont0=\sixrm
	\textfont1=\twelvei  \scriptfont1=\eighti  \scriptscriptfont1=\sixi
	\textfont2=\twelvesy \scriptfont2=\eightsy \scriptscriptfont2=\sixsy
	\textfont3=\twelveex \scriptfont3=\tenex   \scriptscriptfont3=\tenex
	\textfont4=\twelveit \scriptfont4=\eightit \scriptscriptfont4=\sevenit
	\textfont5=\twelvesl
	\textfont6=\twelvebf \scriptfont6=\eightbf \scriptscriptfont6=\sixbf
	\textfont7=\twelvett
	\textfont8=\twelvebi \scriptfont8=\eighti  \scriptscriptfont8=\sixi
	\def\rm{\twelverm\fam=0 }%
	\def\it{\twelveit\fam=4 }%
	\def\sl{\twelvesl\fam=5 }%
	\def\bf{\twelvebf\fam=6 }%
	\def\tt{\twelvett\fam=7 }%
	\def\ss{\twelvess}%
	\def\sc{\twelvesc}%
	\def\bmit{\fam=8 }%
	\rm\setparameters\setbaselines\fi}


\font\fourteenrm=cmr12	scaled\magstep1		
\font\fourteeni=cmmi12	scaled\magstep1		
\font\fourteensy=cmsy10	scaled\magstep2		
\font\fourteenex=cmex10	scaled\magstep2		
\font\fourteenit=cmti12	scaled\magstep1		
\font\fourteensl=cmsl12	scaled\magstep1		
\font\fourteenbf=cmbx12	scaled\magstep1		
\font\fourteentt=cmtt12	scaled\magstep1		
\font\fourteenss=cmss12	scaled\magstep1		
\font\fourteensc=cmcsc10 scaled\magstep2	
\font\fourteenbi=cmmib10 scaled\magstep2	

\skewchar\fourteeni='177
\skewchar\fourteenbi='177
\skewchar\fourteensy='60

\def\fourteenpoint{\ifmmode\err@badsizechange\else
	\textfont0=\fourteenrm \scriptfont0=\tenrm \scriptscriptfont0=\sevenrm
	\textfont1=\fourteeni  \scriptfont1=\teni  \scriptscriptfont1=\seveni
	\textfont2=\fourteensy \scriptfont2=\tensy \scriptscriptfont2=\sevensy
	\textfont3=\fourteenex \scriptfont3=\tenex \scriptscriptfont3=\tenex
	\textfont4=\fourteenit \scriptfont4=\tenit \scriptscriptfont4=\sevenit
	\textfont5=\fourteensl
	\textfont6=\fourteenbf \scriptfont6=\tenbf \scriptscriptfont6=\sevenbf
	\textfont7=\fourteentt
	\textfont8=\fourteenbi \scriptfont8=\tenbi \scriptscriptfont8=\seveni
	\def\rm{\fourteenrm\fam=0 }%
	\def\it{\fourteenit\fam=4 }%
	\def\sl{\fourteensl\fam=5 }%
	\def\bf{\fourteenbf\fam=6 }%
	\def\tt{\fourteentt\fam=7}%
	\def\ss{\fourteenss}%
	\def\sc{\fourteensc}%
	\def\bmit{\fam=8 }%
	\rm\setparameters\setbaselines\fi}


\font\seventeenrm=cmr10 scaled\magstep3		


\newdimen\rp@
\newcount\@basestretchnum
\newskip\@baseskip
\newskip\headskip
\newskip\footskip


\def\setparameters{\rp@=.1em
	\headskip=24\rp@
	\footskip=\headskip
	\delimitershortfall=5\rp@
	\nulldelimiterspace=1.2\rp@
	\scriptspace=0.5\rp@
	\abovedisplayskip=10\rp@ plus3\rp@ minus5\rp@
	\belowdisplayskip=10\rp@ plus3\rp@ minus5\rp@
	\abovedisplayshortskip=5\rp@ plus2\rp@ minus4\rp@
	\belowdisplayshortskip=10\rp@ plus3\rp@ minus5\rp@
	\normallineskip=\rp@
	\lineskip=\normallineskip
	\normallineskiplimit=0pt
	\lineskiplimit=\normallineskiplimit
	\jot=3\rp@
	\setbox0=\hbox{\the\textfont3 B}\p@renwd=\wd0
	\skip\footins=12\rp@ plus3\rp@ minus3\rp@
	\skip\topins=0pt plus0pt minus0pt}


\def\setbaselines{\maxdepth=4\rp@\baselinestretch=\@basestretchnum}


\def\baselinestretch{\afterassignment\@basestretch\@basestretchnum}
\def\@basestretch{%
	\@baseskip=12\rp@ \divide\@baseskip by1000
	\normalbaselineskip=\@basestretchnum\@baseskip
	\baselineskip=\normalbaselineskip
	\bigskipamount=\the\baselineskip
		plus.25\baselineskip minus.25\baselineskip
	\medskipamount=.5\baselineskip
		plus.125\baselineskip minus.125\baselineskip
	\smallskipamount=.25\baselineskip
		plus.0625\baselineskip minus.0625\baselineskip
	\setbox\strutbox=\hbox{\vrule height.708\baselineskip
		depth.292\baselineskip width0pt }}



\def\makeheadline{\vbox to0pt{\baselinestretch=1000
	\vskip-\headskip \vskip1.5pt
	\line{\vbox to\ht\strutbox{}\the\headline}\vss}\nointerlineskip}

\def\makefootline{\baselineskip=\footskip\line{\the\footline}}

\def\big#1{{\hbox{$\left#1\vbox to8.5\rp@ {}\right.\n@space$}}}
\def\Big#1{{\hbox{$\left#1\vbox to11.5\rp@ {}\right.\n@space$}}}
\def\bigg#1{{\hbox{$\left#1\vbox to14.5\rp@ {}\right.\n@space$}}}
\def\Bigg#1{{\hbox{$\left#1\vbox to17.5\rp@ {}\right.\n@space$}}}


\mathchardef\alpha="710B
\mathchardef\beta="710C
\mathchardef\gamma="710D
\mathchardef\delta="710E
\mathchardef\epsilon="710F
\mathchardef\zeta="7110
\mathchardef\eta="7111
\mathchardef\theta="7112
\mathchardef\iota="7113
\mathchardef\kappa="7114
\mathchardef\lambda="7115
\mathchardef\mu="7116
\mathchardef\nu="7117
\mathchardef\xi="7118
\mathchardef\pi="7119
\mathchardef\rho="711A
\mathchardef\sigma="711B
\mathchardef\tau="711C
\mathchardef\upsilon="711D
\mathchardef\phi="711E
\mathchardef\chi="711F
\mathchardef\psi="7120
\mathchardef\omega="7121
\mathchardef\varepsilon="7122
\mathchardef\vartheta="7123
\mathchardef\varpi="7124
\mathchardef\varrho="7125
\mathchardef\varsigma="7126
\mathchardef\varphi="7127
\mathchardef\imath="717B
\mathchardef\jmath="717C
\mathchardef\ell="7160
\mathchardef\wp="717D
\mathchardef\partial="7140
\mathchardef\flat="715B
\mathchardef\natural="715C
\mathchardef\sharp="715D


\def\err@badsizechange{%
	\immediate\write16{--> Size change not allowed in math mode, ignored}}

\baselinestretch=1000
\tenpoint

\catcode`\@=12					
\catcode`\@=11
\expandafter\ifx\csname @iasmacros\endcsname\relax
	\global\let\@iasmacros=\par
\else	\immediate\write16{}
	\immediate\write16{Warning:}
	\immediate\write16{You have tried to input iasmacros more than once.}
	\immediate\write16{}
	\endinput
\fi
\catcode`\@=12


\def\rmb{\seventeenrm}


\def\halfspace{\baselineskip=1.5\normalbaselineskip}
\def\doublespace{\baselineskip=2\normalbaselineskip}


\def\AB{\bigskip\parindent=40pt
        \centerline{\bf ABSTRACT}\medskip\halfspace\narrower}
\def\AE{\bigskip\nonarrower\doublespace}
\def\nonarrower{\advance\leftskip by-\parindent
	\advance\rightskip by-\parindent}


\def\boxit#1{\vbox{\hrule\hbox{\vrule\kern3pt
	\vbox{\kern3pt#1\kern3pt}\kern3pt\vrule}\hrule}}

\def\hence{\leavevmode\hbox{\bf .\raise5.5pt\hbox{.}.} }

\def\dalemb#1#2{{\vbox{\hrule height.#2pt
	\hbox{\vrule width.#2pt height#1pt \kern#1pt \vrule width.#2pt}
	\hrule height.#2pt}}}
\def\gtorder{\mathrel{\raise.3ex\hbox{$>$}\mkern-14mu
             \lower0.6ex\hbox{$\sim$}}}
\def\ltorder{\mathrel{\raise.3ex\hbox{$<$}\mkern-14mu
             \lower0.6ex\hbox{$\sim$}}}

\newdimen\fullhsize
\newbox\leftcolumn
\def\twoup{\hoffset=-.5in \voffset=-.25in
  \hsize=4.75in \fullhsize=10in \vsize=6.9in
  \def\fullline{\hbox to\fullhsize}
  \let\lr=L
  \output={\if L\lr
        \global\setbox\leftcolumn=\columnbox\global\let\lr=R \advancepageno
      \else \doubleformat \global\let\lr=L\fi
    \ifnum\outputpenalty>-20000 \else\dosupereject\fi}
  \def\doubleformat{\shipout\vbox{
    \fullline{\box\leftcolumn\hfil\columnbox}\advancepageno}}
  \def\columnbox{\leftline{\vbox{\makeheadline\pagebody\makefootline}}}
  \tolerance=1000 }
\twelvepoint
\doublespace
{\nopagenumbers{

\rightline{~~~September, 2006}
\bigskip\bigskip
\centerline{\rmb Notes on the Conway-Kochen Twin Argument}
\medskip
\centerline{\it Stephen L. Adler
}
\centerline{\bf Institute for Advanced Study}
\centerline{\bf Princeton, NJ 08540}
\centerline{(adler@ias.edu)}
\medskip
\bigskip\bigskip
\medskip
\bigskip\bigskip
}}
\vfill\eject
\pageno=2
\AB
This is a revision of my original posting, in which I raised objections to
part of the Conway Kochen arument.  I now agree with them that their recent 
reply  answers my original concerns.  In the first part of 
 these notes (identical to the original),  I give a reformulation of the part of  the Conway-Kochen result that closes the contextuality loophole 
in the original Kochen-Specker (KS) theorem.  In the second part (modified 
in this revision) I review my concerns connected with the finite time 
needed to make a measurement, and briefly indicate how Conway and 
Kochen have responded to them.    

\AE
\bigskip\bigskip
\vfill\eject
(1)  {\bf Review of the KS argument.}   Let $n$ be a general spin direction, 
and $H$ a set of hidden variables, which we postulate to determine 
the squared spin of a spin-1 particle in all directions.  That is, 
$$S_n^2=F(H,n)~~~\eqno(1)$$ 
for all $n$ with a fixed set of hidden variables $H$.  To reproduce 
quantum mechanics, KS impose two constraints: 
$$S_n^2 {\rm ~~takes ~only~values~~} 0,1~~~,  \eqno(2a)$$
$${\rm For~every~orthogonal~triple~~} x,y,z, ~~{\rm one~has~~} S_x^2+S_y^2+S_z^2=2
~~~. \eqno(2b)$$
Clearly, the constraints of Eq.~(2a,b) imply that two of $S_x^2,S_y^2,
S_z^2$ are 1, and one is 0.  The KS argument proceeds from constructing 
a set of directions, referred to below as KS directions, for which 
Eqs.~(1a,b) and (2) are in contradiction, implying that there exists no 
function $F(H,n)$ that satisfies the quantum mechanical constraints. 

\noindent 
(2) {\bf Contextuality.}  Let us introduce a restricted notion  
of contextuality as follows.  Let us define a ``relevant parameter'' 
$r$ to be one for which  (i) $r$ must be specified at the start of the 
measurement, in addition to 
$n$, to uniquely describe a spin-measuring apparatus, and 
(ii) $r$ must be given 
at least 2 distinct values to measure $S_n^2$ for all directions in a 
set of KS directions.  We shall assume further that the $r$ value is not  
changed in the course of a single measurement. 
[Example:  A Stern-Gerlach apparatus depends on 
2 directions, the beam axis (say, $x$) and the axis of the inhomogeneous 
magnetic field (say, $z$).  To measure $S_z^2$, the beam axis can be chosen 
to be any direction in a plane perpendicular to $z$. To specify the  
apparatus uniquely, one has to specify the axis $x$, so (i) is satisfied. 
Since no single axis is perpendicular to all of the directions in a 
KS direction set, (ii) is also satisfied.]  Now we can state the contextuality  
loophole:  if $S_n^2=F(H,n,r)$, where the value of $r$ influences whether 
the function $F$ takes the value 0 or 1, then the KS contradiction is avoided. 

\noindent 
(3) {\bf The Conway-Kochen argument to close the loophole.}
Let us introduce the following three assumptions, in addition to the 
constraints of Eq.~(2a,b): 

\noindent {\bf TWIN -- reduced symmetrical version}: Consider a spin-0 state of 
two spin-1 particles in an EPR setup with observer $A$ measuring one particle,  
and observer $B$ measuring the other particle.  Then if $A$  chooses                                        
direction $n$ and measures $S_n^2$,  $B$ must get the same answer $S_n^2$  
on the same direction $n$, irrespective of the distance $d_{AB}$ between 
the two observers.

\noindent{\bf FREE}: Each experimenter can choose the relevant parameter $r$ for 
his/her apparatus and complete a measurement in time $dt$.  (Here,  
for simplicity, $dt$ is 
taken as the minimum and the maximum time for a new measurement.)

\noindent{\bf REL}:  Classical information, such as the value of the relevant 
parameter, propagates with at most a finite signal velocity $c$.  

\leftline{\bf The Conway-Kochen argument}  
To rule out contextuality as defined 
above, proceed as follows: 

\item{~~by TWIN}  $S_n^2=F_A(H,n,r_A) = F_B(H,n,r_B)$
\item{~~by FREE}  $r_A,r_B$ can be set to new values, and a measurement 
performed, in time $dt$. 
\item{~~by REL}  If we take $d_{AB} >> cdt$, then $F_A$ cannot depend on  
$r_B$ and $F_B$ cannot depend on $r_A$, because there is insufficient time 
for a signal to propagate from $A$ to $B$.  Hence we conclude that 

$$F_A(H,n,r_A)=F_B(H,n,r_B) =F_A(H,n)=F_B(H,n)~~~,\eqno(3)$$ 
and we are back to the original KS contradiction.

\noindent(4) {\bf My concern with respect to finite measurement time.}  
Suppose we try to extend the no-go theorem to rule out a contextual 
dependence of the form $F(H,n,i_{\rm past})$, where $i_{\rm past}$ is all 
information in the intersection of the past light cones of the observers 
$A$ and $B$.  Since a  measurement takes a finite time $dt$, and since 
setting up the apparatus for different directions may take differing 
times,  this 
intersection increases (see below) in the course of a measurement, and 
need not be the same for potential measurements in all directions in a 
KS set. 

 To make things concrete, suppose there is a dependence on a signal 
emanating from a distant extra-galactic source, that arrives at identical 
times at $A$ and $B$ (this latter is not an essential restriction).  
Then Eq.~(3) is replaced by 
$$F_A(H,n,S(t))=F_B(H,n,S(t))~~~.\eqno(4)$$
This does not imply that $F_A(H,n,S(t))=F_A(H,n)$, and if the dependence 
on $S(t)$ influences whether $F$ takes the value 0 or 1, then the KS 
contradiction may be  avoided.  
\bigskip
Can this potential problem  be evaded by taking the 
limit $d_{AB} \to \infty$, thereby squeezing the intersection of the 
past cones of $A$ and $B$ back to the infinite past?  The answer is ``no''   
for physically realizable configurations, 
again because of the finiteness of propagation velocities.  In a twin 
experiment, the particles measured by $A$ and $B$ proceed outwards from 
a common initial point.  If the outward velocities were equal to $c$, the 
intersection of the past light cones of $A$ and $B$ would remain a constant, 
equal to the past light cone of the initial point.  However, the 
spin-1 particles used in the KS argument 
must be massive (zero mass spin-1 particles, such as photons, have two 
states of helicity $\pm 1$, but no state of helicity 0).  Hence the 
particles measured by $A$ and $B$ must move apart with velocities less than 
$c$, which implies that the intersection of their past light cones is a 
monotone increasing function of time.  

\noindent(5) {\bf Conway and Kochen's reply}
One can avoid this problem as follows:
  Let $H^{\prime}$ be the information contained within 
the intersection of the past light cones of observers $A$ and $B$ at the 
{\it latest} time potentially involved in a measurement over a  KS set 
of directions.  Since the intersection of the past light cones is monotone 
increasing in time, $H^{\prime}$ represents {\it all} 
information that is used by observers $A$ and $B$, and in particular,   
$H^{\prime}$ contains both the information $H$ and the signal $S(t)$ 
of Eq.~(4).  If one replaces 
$H$ in Eq.~(1) by $H^{\prime}$, one finds, by  
 the original KS argument, that a functional relation 
$$S_n^2=F(H^{\prime},n) ~~~\eqno(5)$$ 
is excluded, which is the result asserted by Conway and Kochen, 
and with which I now agree.    My original concern about this 
argument was that some measurements in the KS set would be at times 
before all of the information in $H^{\prime}$ had arrived, but 
this is in fact  not 
a problem, since the statment that the result of these measurements 
is independent of this later information is not in conflict with 
relativistic causality.

\vfill\eject
\bigskip
\bye